\documentclass[a4paper,12pt]{article}

\usepackage{graphicx}
\usepackage{amsmath}
\usepackage{amssymb}
\usepackage{hyperref}
\usepackage{tabularx}
\newcolumntype{M}{>{$}c<{$}}
\usepackage[matrix,arrow,curve]{xy}

\numberwithin{equation}{section} \numberwithin{figure}{section}
\numberwithin{table}{section}

\def\papertitlepage{\baselineskip 3.5ex\thispagestyle{empty}}
\def\Title#1{\baselineskip 1cm \vspace{1.5cm}%
  \begin{center}{\Large\bf #1}\end{center}\vspace{0.5cm}}
\def\Authors#1{\begin{center}\renewcommand{\thefootnote}{\fnsymbol{footnote}}{\it #1}\end{center}}
\def\Abstract{\vspace{1.0cm}%
  \begin{center}{\large\bf Abstract}\end{center}}

\renewenvironment{thebibliography}{\pagebreak[3]\par\vspace{0.6em}
\begin{flushleft}{\large \bf References}\end{flushleft}
\vspace{-1.0em}

\begin{enumerate}\if@twocolumn\baselineskip=0.6em\itemsep -0.2em
\else\itemsep -0.2em\fi\labelsep 0.1em}{\end{enumerate} }

\DeclareMathDelimiter{\lcolon}{\mathopen}{operators}{"3A}{largesymbols}{"3A}
\DeclareMathDelimiter{\rcolon}{\mathclose}{operators}{"3A}{largesymbols}{"3A}

\def\+{\!\!+\!\!}

\def\dynkin(#1){(#1)}

\def\bra<#1|{\langle#1|}
\def\ket|#1>{|#1\rangle}
\def\braket<#1|#2>{\langle#1|#2\rangle}
\def\llangle{\langle\!\langle}
\def\rrangle{\rangle\!\rangle}
\def\bbra<#1|{\llangle#1|}
\def\kket|#1>{|#1\rrangle}
\def\bbraket<#1|#2>{\llangle#1|#2\rrangle}

\begin{document}
{\papertitlepage \vspace*{0cm} {\hfill
\begin{minipage}{4.2cm}
IF-USP 2010\par\noindent November, 2010
\end{minipage}}
\Title{Comments on regularization of identity based solutions in
string field theory}
\Authors{{\sc E.~Aldo~Arroyo${}$\footnote{\tt
aldohep@fma.if.usp.br}}
\\
Instituto de F\'{i}sica, Universidade de S\~{a}o Paulo \\[-2ex]
C.P. 66.318 CEP 05315-970, S\~{a}o Paulo, SP, Brasil ${}$ }
} 

\vskip-\baselineskip
{\baselineskip .5cm \Abstract We analyze the consistency of the
recently proposed regularization of an identity based solution in
open bosonic string field theory. We show that the equation of
motion is satisfied when it is contracted with the regularized
solution itself. Additionally, we propose a similar regularization
of an identity based solution in the modified cubic superstring
field theory.
 }
\newpage
\setcounter{footnote}{0}
\tableofcontents

\section{Introduction}
In a previous work \cite{Arroyo:2010fq}, we have shown a
prescription for computing identity based solutions in cubic-like
string field theories
\cite{Witten:1985cc,Preitschopf:1989fc,Arefeva:1989cp}. Although
these identity based solutions provide ambiguous result for the
value of the vacuum energy \cite{Kishimoto:2009nd}, we noticed
that the tractable Erler-Schnabl's solution of open bosonic string
field theory \cite{Erler:2009uj} is related by a gauge
transformation to a solution which is based on the identity string
field. Moreover, we proved that the same is true in the case of
the modified cubic superstring field theory, namely the regular
solution of Gorbachev \cite{Gorbachev:2010zz} is related by a
gauge transformation to an identity based solution.

After performing the gauge transformation, the resulting
Erler-Schnabl-type solutions were used to unambiguously compute
the value of the vacuum energy. Nevertheless, it would be
interesting to evaluate directly the vacuum energy using the
identity based solutions, this kind of computation should be
possible provide that we can find a consistent regularization
scheme. Recently a proposal for regularizing an identity based
solution in open bosonic string field theory was developed in
\cite{Zeze:2010sr}.

The regularized solution $\Psi_\lambda$ was obtained by
considering one-parameter families of classical solutions
\begin{eqnarray}
\label{regularsol1intro} \Psi_\lambda = U_\lambda Q U_\lambda^{-1}
+ U_\lambda \Psi_I U_\lambda^{-1} \, ,
\end{eqnarray}
where $\Psi_I=c(1-K)$ is the identity based solution found in
\cite{Arroyo:2010fq} and
\begin{eqnarray}
\label{gauge01intro} U_\lambda=1+\lambda cBK\, , \;\;\;
U_\lambda^{-1}=1-\lambda cBK \frac{1}{1+\lambda K}
\end{eqnarray}
is an element of the gauge transformation
\cite{Arroyo:2010fq,Zeze:2010sr}. It has been shown that the
resulting regularized solution $\Psi_\lambda$
\begin{eqnarray}
\label{regularsol2intro} \Psi_\lambda = (c+\lambda c KB c)
\frac{1+(\lambda-1)K}{1+\lambda K} \,,
\end{eqnarray}
correctly reproduces the value of the kinetic term
\begin{eqnarray}
\label{accionI1} \langle  \Psi_\lambda,Q \Psi_\lambda
\rangle=-\frac{3\;}{\pi^2} \, ,
\end{eqnarray}
and therefore 'assuming the equation of motion', the right value
of the vacuum energy was reproduced. We have put assuming the
equation of motion in quotation marks, since it remains as an
important question if the regularization is consistent with the
assumption that the equation of motion is satisfied when it is
contracted with the solution itself, namely
\begin{eqnarray}
\label{accionI2}  \langle\Psi_\lambda,Q \Psi_\lambda
\rangle+\langle \Psi_\lambda,\Psi_\lambda*\Psi_\lambda \rangle=0
\, .
\end{eqnarray}

From previous experiences in the past
\cite{Takahashi:2002ez,Kishimoto:2002xi,Okawa:2003cm,Okawa:2006vm,Fuchs:2006hw,Arroyo:2009ec},
it is clear that there is a subtlety about this assumption because
in general the solution is usually outside the Fock space
\cite{Okawa:2006vm}. For instance the twisted butterfly state
\cite{Gaiotto:2001ji,Schnabl:2002ff,Gaiotto:2002kf} in vacuum
string field theory \cite{Rastelli:2001jb} solves the equation of
motion when contracted with any state in the Fock space, but it
does not satisfy the equation of motion when contracted with the
solution itself \cite{Okawa:2003cm}. Therefore, for the case of
the regularized solution $\Psi_\lambda$, it is important to test
the validity of the assumption (\ref{accionI2}) and for this goal
it is necessary to evaluate the cubic term of the string field
theory action for the regularized solution $\Psi_\lambda$ to check
if the right value is reproduced \footnote{Let us point out that a
similar test of consistency was performed by Okawa
\cite{Okawa:2006vm}, Fuchs, Kroyter \cite{Fuchs:2006hw} and Arroyo
\cite{Arroyo:2009ec} for the original Schnabl's solution
\cite{Schnabl:2005gv}.}
\begin{eqnarray}
\label{cubicI1}  \langle \Psi_\lambda,\Psi_\lambda*\Psi_\lambda
\rangle=\frac{3\;}{\pi^2} \,.
\end{eqnarray}

In this paper we compute the analytical value of the kinetic
(\ref{accionI1}) and the cubic term (\ref{cubicI1}) of the string
field theory action for the regularized solution $\Psi_\lambda$
and we show that the assumption of the equation of motion
(\ref{accionI2}) was nevertheless correct\footnote{Let us point
out that the analytical value of the kinetic term (\ref{accionI1})
was already calculated in reference \cite{Zeze:2010sr}.
Nevertheless, for completeness reasons, in this paper we are going
to review this computation.}. In addition to our analytic results,
using Pad\'{e} approximants \cite{Arroyo:2009ec} we numerically
test equations (\ref{accionI1}) and (\ref{cubicI1}) for the
particular values $\lambda \rightarrow 0$ and $\lambda \rightarrow
1$ which correspond to the identity based and Erler-Schnabl's
solution respectively. We would like to comment that there are
cases where, naively, a one-parameter family of gauge
transformations connects two solutions that cannot be
gauge-equivalent (see for instance references
\cite{Okawa:2006vm,Fuchs:2006hw,Ellwood:2009zf,Takahashi:2007du}).
What can go wrong is that at some particular values of the
parameter the gauge transformation becomes singular
\cite{Ellwood:2009zf}. So at this point it is interesting to ask:
whether this problem affects, or not, the gauge transformations
(\ref{regularsol1intro}) and (\ref{gauge01intro}). Do these gauge
transformations become singular at some particular values of the
parameter $\lambda$? We will show that the gauge transformations
(\ref{regularsol1intro}) and (\ref{gauge01intro}) are well-defined
for all values of the parameter $\lambda$ belonging to the
interval $[0,+\infty)$. These results provide a non trivial
evidence for the consistency of the regularization proposed in
\cite{Zeze:2010sr}.

Finally, we propose a similar regularization for an identity based
solution in the modified cubic superstring field theory and, as in
the bosonic case, we show that the regularized solution
consistently reproduces the right value for the kinetic and cubic
term and consequently for the vacuum energy. Our results show
explicitly that how seemingly trivial identity based solutions, in
open bosonic string field theory as well as in the modified cubic
superstring field theory, can be consistently regularized to
obtain well behaved solutions which precisely represent to the
tachyon vacuum. Certainly it would be very interesting to extend
these results to the case of the non-polynomial Berkovits WZW-type
superstring field theory \cite{Berkovits:1995ab}.

This paper is organized as follows. In section 2, we review the
proposal for regularizing an identity based solution in open
bosonic string field theory. We evaluate the kinetic and cubic
term of the string field theory action for the regularized
solution. It turns out that the value of the cubic term is
correctly reproduced and therefore we prove the statement that the
equation of motion is satisfied when it is contracted with the
regularized solution itself. In section 3, we regularize an
identity based solution in the modified cubic superstring field
theory. As in the bosonic case, in order to prove the validity of
the assumption that the equation of motion is satisfied when it is
contracted with the regularized solution itself, we evaluate the
kinetic and cubic term. In section 4, a summary and further
directions of exploration are given. The appendix A is provided
for explaining some details related to the computation of
correlation functions. The appendix B is devoted to some explicit
Pad\'{e} approximants computations.

\section{Regularization of identity based solution in open bosonic string field theory}

As derived in \cite{Arroyo:2010fq} using the methods of
\cite{Erler:2006hw,Erler:2006ww}, an identity based solution in
open bosonic string field theory is given by
\begin{eqnarray}
\label{identity01} \Psi_I=c(1-K) \,
\end{eqnarray}
where the basic string fields $c$ and $K$ (together with $B$) can
be written, using the operator representation
\cite{Schnabl:2005gv}, as follows
\begin{eqnarray}
\label{K01} K &\rightarrow& \frac{1}{2} \hat{\mathcal{L}}
U_{1}^\dag U_{1} |0\rangle \, ,
\\
\label{B01} B &\rightarrow& \frac{1}{2} \hat{\mathcal{B}}
U_{1}^\dag U_{1} |0\rangle \, ,
\\
\label{c01} c &\rightarrow&   U_{1}^\dag U_{1} \tilde c
(0)|0\rangle \, .
\end{eqnarray}

The operators $\hat{\mathcal{L}}$, $\hat{\mathcal{B}}$ and $\tilde
c(0)$ are defined in the sliver frame \cite{Erler:2009uj}
\footnote{Remember that a point in the upper half plane $z$ is
mapped to a point in the sliver frame $\tilde z$ via the conformal
mapping $\tilde z= \frac{2}{\pi}\arctan z $. Note that we are
using the convention of \cite{Erler:2009uj} for the conformal
mapping.}, and they are related to the worldsheet energy-momentum
tensor, the $b$ and $c$ ghosts fields respectively, for instance
\begin{eqnarray}
\label{Lhat01} \hat{\mathcal{L}} &\equiv& \mathcal{L}_{0} +
\mathcal{L}^{\dag}_0 = \oint \frac{d z}{2 \pi i} (1+z^{2})
(\arctan z+\text{arccot} z) \,
T(z) \, , \\
\label{Bhat01} \hat{\mathcal{B}} &\equiv& \mathcal{B}_{0} +
\mathcal{B}^{\dag}_0 = \oint \frac{d z}{2 \pi i} (1+z^{2})
(\arctan z+\text{arccot} z) \, b(z) \, ,
\end{eqnarray}
while the operator $U_{1}^\dag U_{1}$ in general is given by
$U^\dag_r U_r = e^{\frac{2-r}{2} \hat{\mathcal{L}}}$, so we have
chosen $r=1$, note that the string field $U_{1}^\dag U_{1}
|0\rangle$ represents to the identity string field $1 \rightarrow
U_{1}^\dag U_{1} |0\rangle$
\cite{Okawa:2006vm,Schnabl:2005gv,Erler:2006hw,Erler:2006ww}.

Using the operator representation (\ref{K01})-(\ref{c01}) of the
string fields $K$, $B$ and $c$, we can show that these fields
satisfy the algebraic relations
\begin{eqnarray}
\label{eq2} \{B,c\}=1\, , \;\;\;\;\;\;\; [B,K]=0 \, ,
\;\;\;\;\;\;\; B^2=c^2=0 \, ,
\end{eqnarray}
and have the following BRST variations
\begin{eqnarray}
\label{eq3} QK=0 \, , \;\;\;\;\;\; QB=K \, , \;\;\;\;\;\; Qc=cKc
\, .
\end{eqnarray}

As it is shown in \cite{Arroyo:2010fq} the direct evaluation of
the vacuum energy using the identity based solution
(\ref{identity01}) brings ambiguous result. This phenomenon, as it
was noted in \cite{Zeze:2010sr}, is due to the fact that a naive
evaluation of the classical action in terms of CFT methods tends
to be indefinite since it corresponds to a correlator on vanishing
strip. Recently this problem was overcome and a proposal for
regularizing the identity based solution (\ref{identity01}) has
been developed in \cite{Zeze:2010sr}.

The regularized solution $\Psi_\lambda$ is obtained by considering
one-parameter families of classical solutions
\begin{eqnarray}
\label{regularsol1} \Psi_\lambda = U_\lambda Q U_\lambda^{-1} +
U_\lambda \Psi_I U_\lambda^{-1} \, ,
\end{eqnarray}
where $\Psi_I$ is the identity based solution (\ref{identity01})
and
\begin{eqnarray}
\label{gauge01} U_\lambda=1+\lambda cBK\, , \;\;\;
U_\lambda^{-1}=1-\lambda cBK \frac{1}{1+\lambda K}
\end{eqnarray}
is an element of the gauge transformation
\cite{Arroyo:2010fq,Zeze:2010sr}. Using (\ref{identity01}),
(\ref{regularsol1}) and (\ref{gauge01}), it is almost easy to
derive the following regularized solution
\begin{eqnarray}
\label{regularsol2} \Psi_\lambda = c(1+\lambda K) Bc
\frac{1+(\lambda-1)K}{1+\lambda K} \,.
\end{eqnarray}

Note that this regularized solution interpolates between the
identity based solution (\ref{identity01}) which corresponds to
the case $\lambda\rightarrow 0$, and the Erler-Schnabl's solution
\cite{Erler:2009uj} which corresponds to the case
$\lambda\rightarrow 1$. In the next subsection we are going to
evaluate the kinetic term for the regularized solution, and it
will be shown that its value does not depend on the parameter
$\lambda$.

\subsection{The kinetic term}
In this subsection, we are going to evaluate the kinetic term of
the string field theory action for the regularized solution
$\Psi_\lambda$ \footnote{Let us point out that, in the bosonic
case, the analytical calculation of the kinetic term was already
performed in \cite{Zeze:2010sr}. Nevertheless, for completeness
reasons, in this subsection we are going to review this
computation.}
\begin{eqnarray}
\label{kinetic001} \langle  \Psi_\lambda,Q \Psi_\lambda \rangle
\,.
\end{eqnarray}

Since the regularized solution (\ref{regularsol2}) can be written
as an expression containing an exact BRST term
\begin{eqnarray}
\label{regularsol3} \Psi_\lambda = c
\frac{1+(\lambda-1)K}{1+\lambda K}+ Q \Big\{ \lambda Bc
\frac{1+(\lambda-1)K}{1+\lambda K} \Big\} \,,
\end{eqnarray}
the computation of the kinetic term (\ref{kinetic001}) can be
simplified to the evaluation of the following correlator
\begin{eqnarray}
\label{kinetic002} \langle  \Psi_\lambda,Q \Psi_\lambda \rangle
&=& \langle  c \frac{1+(\lambda-1)K}{1+\lambda K} cKc
\frac{1+(\lambda-1)K}{1+\lambda K} \rangle \nonumber \\
&=& \int_0^{\infty}\int_0^{\infty} dt_1dt_2 e^{-t_1-t_2} \langle c
(1+(\lambda-1)K)\Omega^{\lambda t_1} cKc
(1+(\lambda-1)K)\Omega^{\lambda t_2} \rangle \nonumber \\
&=&\int_0^{\infty}\int_0^{\infty} dt_1dt_2 e^{-t_1-t_2}
\Big(1+\frac{1-\lambda}{\lambda}
\partial_{t_1}\Big)\Big(1+\frac{1-\lambda}{\lambda}
\partial_{t_2}\Big)\langle c \Omega^{\lambda t_1} cKc
\Omega^{\lambda t_2} \rangle \, .\nonumber \\
\end{eqnarray}

Using the expression for the correlator $\langle c \Omega^{\lambda
t_1} cKc \Omega^{\lambda t_2} \rangle$ (given in the appendix),
applying the change of variables as in \cite{Erler:2009uj} $t_1
\rightarrow u v$, $t_2 \rightarrow u (1 - v)$ and performing the
$v$ integral, we get from (\ref{kinetic002})
\begin{eqnarray}
\label{kinetic003} \langle  \Psi_\lambda,Q \Psi_\lambda \rangle
&=& -\frac{1}{2 \pi^2} \int_0^{\infty} du e^{-u} \big[ 6 (\lambda
-1)^2 u -6 (\lambda -1) \lambda u^2 +\lambda ^2 u^3\big] \nonumber
\\
&=&-\frac{3}{\pi ^2} \, .
\end{eqnarray}
Therefore, as it was previously commented, the value of the
kinetic term does not depend on the parameter $\lambda$. At this
stage, we can safely take the limit $\lambda \rightarrow 0$ which
corresponds to the identity based solution.

If we assume the validity of the equation of motion when
contracted with the regularized solution itself, it is clear that
the value of the vacuum energy can be correctly reproduced
\cite{Zeze:2010sr}. Nevertheless, there is a subtlety about this
assumption because in general the solution is usually outside the
Fock space \cite{Okawa:2006vm}. Therefore it is crucially
important to know whether or not the equation of motion is
satisfied when it is contracted with the regularized solution
itself. To prove the correctness of this statement, it is
necessary to evaluate the cubic term of the string field theory
action.

\subsection{The cubic term}
In this subsection, we are going to evaluate the cubic term of the
string field theory action for the regularized solution
\begin{eqnarray}
\label{cubic001x} \langle  \Psi_\lambda,\Psi_\lambda *
\Psi_\lambda \rangle \,.
\end{eqnarray}

Since the regularized solution (\ref{regularsol2}) can be written
as an expression containing two terms
\begin{eqnarray}
\label{regularsolx1} \Psi_\lambda &=& \Psi_1 + \Psi_2 \, , \\
\label{regularsolx2} \Psi_1&=& c \frac{1+(\lambda-1)K}{1+\lambda
K} \, , \\
\label{regularsolx3} \Psi_2&=&  \lambda cBKc
\frac{1+(\lambda-1)K}{1+\lambda K} \, ,
\end{eqnarray}
the calculation of the cubic term (\ref{cubic001x}) can be reduced
to the evaluation of the following correlators
\begin{eqnarray}
\label{cubic002x} \langle  \Psi_\lambda, \Psi_\lambda *
\Psi_\lambda\rangle &=& \langle \Psi_1, \Psi_1 * \Psi_1\rangle  +
3 \langle \Psi_2, \Psi_2 * \Psi_1\rangle + 3 \langle \Psi_2,
\Psi_1 * \Psi_1\rangle + \langle \Psi_2, \Psi_2 * \Psi_2\rangle \,
, \nonumber \\
\end{eqnarray}
each term on the right hand side (RHS) of (\ref{cubic002x}) is
given by
\begin{eqnarray}
\label{cubic003x1} \langle \Psi_1, \Psi_1 * \Psi_1\rangle &=&
\int_0^{\infty}\int_0^{\infty}\int_0^{\infty} dt_1dt_2dt_3
e^{-t_1-t_2-t_3} \mathcal{D}_1 \mathcal{D}_2 \mathcal{D}_3 \langle
c \Omega^{\lambda t_1} c \Omega^{\lambda t_2} c \Omega^{\lambda
t_3} \rangle  \, , \\
\label{cubic003x2} \langle \Psi_2, \Psi_2 * \Psi_1\rangle &=&
\lambda^2 \int_0^{\infty}\int_0^{\infty}\int_0^{\infty}
dt_1dt_2dt_3 e^{-t_1-t_2-t_3} \mathcal{D}_1 \mathcal{D}_2
\mathcal{D}_3 \langle cBKc \Omega^{\lambda t_1} cBKc
\Omega^{\lambda t_2} c \Omega^{\lambda t_3} \rangle \nonumber \, ,
\\\\
\label{cubic003x3} \langle \Psi_2, \Psi_1 * \Psi_1\rangle &=&
\lambda \int_0^{\infty}\int_0^{\infty}\int_0^{\infty} dt_1dt_2dt_3
e^{-t_1-t_2-t_3} \mathcal{D}_1 \mathcal{D}_2 \mathcal{D}_3 \langle
cBKc \Omega^{\lambda t_1} c \Omega^{\lambda t_2} c \Omega^{\lambda
t_3} \rangle \nonumber \, , \\
\end{eqnarray}
where the differential operators $\mathcal{D}_1$, $\mathcal{D}_2$
and $\mathcal{D}_3$ are defined as
\begin{eqnarray}
\label{diff1} \mathcal{D}_i &\equiv& 1+\frac{1-\lambda}{\lambda}
\partial_{t_i}\; , \;\;\;\;\; i=1,2,3\, .
\end{eqnarray}
Since the string field $\Psi_2$ defined in (\ref{regularsolx3})
can be written as an exact BRST term (\ref{regularsol3}), the last
term on the RHS of (\ref{cubic002x}) gives vanishing result.

Using the expression for the correlators $\langle c
\Omega^{\lambda t_1} c \Omega^{\lambda t_2} c \Omega^{\lambda t_3}
\rangle$, $\langle cBKc \Omega^{\lambda t_1} cBKc \Omega^{\lambda
t_2} c \Omega^{\lambda t_3} \rangle$ and $\langle cBKc
\Omega^{\lambda t_1} c \Omega^{\lambda t_2} c \Omega^{\lambda t_3}
\rangle$ (given in the appendix) into equations
(\ref{cubic003x1})$-$(\ref{cubic003x3}), applying the change of
variables as in \cite{Zeze:2010jv} $t_1 \rightarrow u v_1$, $t_2
\rightarrow  u v_2$, $t_3 \rightarrow  u (1 - v_1 - v_2)$ and
performing the $v_1,v_2$ integral, we obtain from
(\ref{cubic002x})
\begin{align}
\label{cubic0y1} \langle  \Psi_\lambda, \Psi_\lambda *
\Psi_\lambda\rangle &= \frac{1}{8 \pi ^4} \int_0^{\infty} du
e^{-u} \Big[ \; 24 (-15+\pi ^2) (\lambda -1)^3  -24 (-15+\pi
^2) (\lambda -1)^2 (5 \lambda -2) u \nonumber \\
&\;\;\;\; + 36 (\lambda -1) (3 \pi ^2 \lambda ^2-50 \lambda ^2-2
\pi ^2 \lambda +40 \lambda -5) u^2 + \lambda ^2 (2 \pi ^2 \lambda
-75 \lambda
+45) u^4 \nonumber \\
&\;\;\;\; -4 \lambda  (7 \pi ^2 \lambda ^2-150 \lambda ^2-6 \pi ^2
\lambda +180 \lambda -45)u^3 + 3 \lambda ^3 u^5 \; \Big] \nonumber
\\
&=\frac{3}{\pi ^2} \, .
\end{align}

We see that the value of the cubic term does not depend on the
parameter $\lambda$. This result (\ref{cubic0y1}) proves the
statement that the equation of motion is satisfied when it is
contracted with the regularized solution itself.

\subsection{$\mathcal{L}_0$ level expansion and Pad\'{e} approximants}

Although we have an analytic result for the value of the kinetic
(\ref{kinetic003}) and cubic term (\ref{cubic0y1}), we would like
to confirm our calculation by using the $\mathcal{L}_0$ level
expansion of the solution. A numerical method based on the
$\mathcal{L}_0$ level expansion of the solution was developed in
references \cite{Erler:2009uj,Arroyo:2009ec,Schnabl:2005gv}, where
the vacuum energy for the original Schnabl's solution
\cite{Schnabl:2005gv} was represented as a formal sum of an
asymptotic series which was resumed using Pad\'{e} approximants.
In this subsection, using this $\mathcal{L}_0$ level expansion
scheme, we are going to numerically test equations
(\ref{kinetic003}) and (\ref{cubic0y1}) for the particular values
$\lambda \rightarrow 0$ and $\lambda \rightarrow 1$ which
correspond to the identity based (\ref{identity01}) and
Erler-Schnabl's solution \cite{Erler:2009uj} respectively.

In order to evaluate the kinetic and cubic term of the string
field theory action in the $\mathcal{L}_0$ level expansion scheme,
let us write the regularized solution (\ref{regularsol2}) in terms
of $\mathcal{L}_0$ eigenstates
\begin{align}
\label{L0Bos} \Psi_\lambda =& \sum_{n,p} f_{n,p}(\lambda) ({\cal
L}_0 + {\cal L}_0^\dagger)^n \tilde c_p |0\rangle  + \sum_{n,p,q}
f_{n,p,q}(\lambda) ({\cal B}_0 + {\cal B}_0^\dagger) ({\cal L}_0 +
{\cal
L}_0^\dagger)^n \tilde c_p \tilde c_q  |0\rangle \, , \\
f_{n, p}(\lambda) =& 2^{-n} \frac{(\lambda -1)}{\lambda}
\left(\frac{1}{
   n!}+\frac{\lambda}{(n-1)!}\right) \delta _{1,p} \nonumber \\
  &+\frac{2^{-n+p-1}}{\lambda } \sum_{k=0}^{n} \frac{(-1)^{n-k} \lambda ^{n-k-p+1} (n-k-p+1)!}{k!
   (n-k)!} \nonumber \\&+ 2^{-n+p-1} \sum_{k=0}^{n-1} \frac{(-1)^{n-k-1} \lambda ^{n-k-p} (n-k-p)!}{k! (n-k-1)!} \, , \\
f_{n, p, q}(\lambda) =& \frac{2^{-n-2} (1-\lambda )}{n!}
\big(\delta _{0,q} \delta _{1,p}-\delta _{0,p} \delta _{1,q} \big)
\nonumber \\
&+2^{-n+p+q-3} (q-p) \sum_{k=0}^{n}\frac{(-1)^{n-k} \lambda
^{n-k-p-q+1} (n-k-p-q+1)!}{k! (n-k)!}\, .
\end{align}

As it is described in
\cite{Erler:2009uj,Arroyo:2009ec,Schnabl:2005gv}, we start by
replacing the solution $\Psi_\lambda$ with $z^{\mathcal{L}_0}
\Psi_\lambda$ in the $\mathcal{L}_0$ level truncation scheme, so
that states in the $\mathcal{L}_0$ level expansion of the solution
will acquire different integer powers of $z$ at different levels.
As we are going to see, the parameter $z$ is needed because we
need to express the kinetic and cubic term as a formal power
series expansion if we want to use Pad\'{e} approximants. After
doing our calculations, we will simply set $z = 1$. Plugging the
$\mathcal{L}_0$ level expansion of the regularized solution
(\ref{L0Bos}) into the kinetic term we obtain
\begin{align}
\label{kineticPa} \langle \Psi_\lambda,
z^{\mathcal{L}_0^{\dagger}} Q z^{\mathcal{L}_0} \Psi_\lambda
\rangle =&-\frac{4}{\pi ^2 z^2}+\frac{(\pi ^2-4) (2 \lambda -1)}{2
\pi ^2} + \frac{\pi ^2 (\lambda ^2-2 \lambda^3 )}{8}  z^2 +
\frac{\pi ^2  (5 \lambda^4-4 \lambda^3 +\lambda ^2)}{4} z^3 \nonumber \\
&+ \frac{ \pi ^2 (2 \pi ^2 \lambda ^5-180 \lambda ^5-\pi ^2
\lambda ^4+174 \lambda ^4-72 \lambda ^3+12
   \lambda ^2)}{32} z^4 \nonumber \\ &+ \frac{\pi^2 (-7 \pi ^2 \lambda ^6
   +210 \lambda ^6+5 \pi ^2 \lambda ^5-222 \lambda ^5-\pi ^2 \lambda ^4+114 \lambda ^4-32 \lambda
   ^3+4 \lambda ^2)}{8} z^5 \nonumber \\ &+\cdots \, .
\end{align}

Given this formal power series expansion (\ref{kineticPa}) of the
kinetic term, we are going to resum the series using Pad\'{e}
approximants for some particular values of the parameter
$\lambda$. Basically the numerical method based on Pad\'{e}
approximants tell us to match the power series expansion
coefficients of a given rational function $P_{2+N}^{M}(z)$ with
those of the kinetic term (\ref{kineticPa}). The details of these
computations can be found in appendix B.

The result of our calculations is summarized in table
\ref{results1}. In the first column we show the normalized value
of the kinetic term $\frac{\pi^2}{3}\langle \Psi_\lambda,
z^{\mathcal{L}_0^{\dagger}} Q z^{\mathcal{L}_0} \Psi_\lambda
\rangle$ for the particular value of the parameter $\lambda
\rightarrow 0$ which corresponds to the identity based solution.
In the second column we show the normalized value of the kinetic
term for the particular value of the parameter $\lambda
\rightarrow 1$ which corresponds to the Erler-Schnabl's solution
\footnote{Actually the numerical computation of the kinetic term
using Pad\'{e} approximants for the particular value of the
parameter $\lambda \rightarrow 1$ was already performed in
reference \cite{Erler:2009uj}.}. As we can see from table
\ref{results1} the value of the kinetic term computed numerically
using Pad\'{e} approximants nicely confirm the analytic result
(\ref{kinetic003}).

\begin{table}[ht]
\caption{The Pad\'{e} approximation for the normalized value of
the kinetic term $\frac{\pi^2}{3}\langle \Psi_\lambda,
z^{\mathcal{L}_0^{\dagger}} Q z^{\mathcal{L}_0} \Psi_\lambda
\rangle$ evaluated at $z=1$. The first column corresponds to
$P_{2+n}^{n}$ Pad\'{e} approximation for the value of the
parameter $\lambda \rightarrow 0$, while in the second column we
show the case $\lambda \rightarrow 1$. The label $n$ corresponds
to the power of $z$ in the series (\ref{kineticPa}). At each stage
of our computations we truncate the series up to the order
$z^{2n-2}$.} \centering
\begin{tabular}{|c|c|c|}
\hline
  &  $P^{n}_{2+n}(\lambda \rightarrow 0)$ Pad\'{e} approximation   &  $P^{n}_{2+n}(\lambda \rightarrow 1)$ Pad\'{e} approximation     \\
    \hline
$n=0$   &  $-1.333333333333$ &  $-1.333333333333$ \\
\hline  $n=2$ & $-2.311600733514$  & $-1.143337106188$  \\
\hline $n=4$   & $-1.494531194598$ & $-0.898882661597$  \\
\hline $n=6$ & $-1.077721474044$ & $-1.042410506615$ \\
\hline $n=8$  & $-1.606155676411$ & $-0.996478424643$  \\
\hline $n=10$ & $-0.951428845113$ & $-0.995773031227$   \\
\hline $n=12$ & $-1.008770444432$  & $-0.999000106158$  \\
\hline $n=14$  & $-1.001300960887$  & $-0.999417017099$  \\
\hline $n=16$ & $-1.005286583496$  & $-0.999198792613$  \\
\hline
\end{tabular}
\label{results1}
\end{table}

To evaluate the cubic term using its formal power series expansion
in $z$ by means of Pad\'{e} approximants, we follow the same steps
developed in the case of the kinetic term. First we plug the
$\mathcal{L}_0$ level expansion of the regularized solution
(\ref{L0Bos}) into the expression for the cubic term
\begin{align}
\label{expansion1} \langle \Psi_\lambda ,z^{{\cal
L}_0^\dag}(z^{{\cal L}_0}\Psi_\lambda)*(z^{{\cal
L}_0}\Psi_\lambda) \rangle  =& \frac{81 \sqrt{3}}{8
\pi^3}\frac{1}{z^3}+\frac{27 \left(-3 \sqrt{3}+\pi \right) \lambda }{4 \pi ^3 z^2}  \nonumber \\
&+ \frac{4 \sqrt{3} \pi ^2 \left(2 \lambda ^2-6 \lambda
+3\right)+27
   \sqrt{3} \left(2 \lambda ^2+6 \lambda -3\right)-36 \pi \lambda ^2}{8 \pi ^3 z} \nonumber \\
&+ \frac{\pi ^3 (8-12 \lambda )-36 \sqrt{3} \pi ^2 (\lambda
-1)^2+81 \pi  \lambda  (2
   \lambda -1)+81 \sqrt{3} \left(1-3 \lambda ^2\right)}{36 \pi ^3}   \nonumber \\
&+ \frac{2 \pi  \lambda  (3 \lambda -2)}{27 \sqrt{3}} z+\frac{4
\pi  \lambda  \left(3 (-6 \sqrt{3}+\pi ) \lambda ^2+(24 \sqrt{3}-2
   \pi ) \lambda -9 \sqrt{3}\right)}{729} z^2  \nonumber \\
& + \, \cdots \, .
\end{align}

Given the formal power series expansion (\ref{expansion1}), we are
able to evaluate the cubic term using Pad\'{e} approximants. We
match the power series expansion coefficients of a given rational
function $P_{3+N}^M(z)$ with those of the cubic term
(\ref{expansion1}). The result of our computations is summarized
in table \ref{results2}. In the first column we show the
normalized value of the cubic term $\frac{\pi^2}{3}\langle
\Psi_\lambda ,z^{{\cal L}_0^\dag}(z^{{\cal
L}_0}\Psi_\lambda)*(z^{{\cal L}_0}\Psi_\lambda) \rangle$ for the
particular value of the parameter $\lambda \rightarrow 0$ which
corresponds to the identity based solution. In the second column
we show the normalized value of the cubic term for the particular
value of the parameter $\lambda \rightarrow 1$ which corresponds
to the Erler-Schnabl's solution. As we can see from table
\ref{results2} the value of the cubic term computed numerically
using Pad\'{e} approximants nicely confirm the analytic result
(\ref{cubic0y1}).

\begin{table}[ht]
\caption{The Pad\'{e} approximation for the normalized value of
the cubic term $\frac{\pi^2}{3}\langle \Psi_\lambda ,z^{{\cal
L}_0^\dag}(z^{{\cal L}_0}\Psi_\lambda)*(z^{{\cal
L}_0}\Psi_\lambda) \rangle$ evaluated at $z=1$. The first column
corresponds to $P^{n/2}_{3+n/2}$ Pad\'{e} approximation for the
value of the parameter $\lambda \rightarrow 0$, while in the
second column we show the case $\lambda \rightarrow 1$. The label
$n$ corresponds to the power of $z$ in the series
(\ref{expansion1}). At each stage of our computations we truncate
the series up to the order $z^{n-3}$.} \centering
\begin{tabular}{|c|c|c|}
\hline
  &  $P^{n/2}_{3+n/2}(\lambda \rightarrow 0)$ Pad\'{e} approximation   &  $P^{n/2}_{3+n/2}(\lambda \rightarrow 1)$ Pad\'{e} approximation     \\
    \hline
$n=0$   &  $1.860735022048$ &  $1.860735022048$ \\
\hline  $n=2$ & $1.860735022048$  & $0.860998808763$  \\
\hline $n=4$   & $2.221490574460$ & $0.839819361621$  \\
\hline $n=6$ & $2.051478161167$ & $0.883381212050$ \\
\hline $n=8$  & $0.893773019606$ & $0.967208479640$  \\
\hline $n=10$ & $1.088076700691$ & $0.983453704434$   \\
\hline $n=12$ & $1.038107274783$  & $0.964647638538$  \\
\hline $n=14$  & $1.005610433784$  & $0.995649817299$  \\
\hline $n=16$ & $1.003669862242$  & $0.997027139055$  \\
\hline
\end{tabular}
\label{results2}
\end{table}

As it was mentioned in the introduction section, at this point we
would like to argue that the gauge transformations
(\ref{regularsol1intro}) and (\ref{gauge01intro}) are well-defined
for all values of the parameter $\lambda$ belonging to the
interval $[0,+\infty)$. In order to develop the arguments to prove
this statement, let us review the discussion given for the case of
the original Schnabl's solution \cite{Schnabl:2005gv}. It is known
that Schnabl's solution can be written as the limit, $\lambda
\rightarrow 1$, of the following pure gauge form
\cite{Okawa:2006vm,Fuchs:2006hw,Ellwood:2009zf}
\begin{eqnarray}
\label{puregauge1} \Psi^{\text{S}}_\lambda = \Gamma_\lambda Q
\Gamma^{-1}_\lambda \, ,
\end{eqnarray}
where S stands for Schnabl's solution and
\begin{eqnarray}
\label{puregauge2} \Gamma_\lambda = 1-\lambda \Phi \, ,
\;\;\;\;\;\; \Gamma^{-1}_\lambda = \frac{1}{ 1-\lambda \Phi}
\end{eqnarray}
with
\begin{eqnarray}
\label{puregauge3} \Phi =B_1^{L} \tilde c_1|0\rangle.
\end{eqnarray}

When $\lambda<1$ the string fields $\Psi^{\text{S}}_\lambda$ and
$\Gamma_\lambda^{-1}$ are well-defined in the level-expansion and
the solution $\Psi^{\text{S}}_\lambda$ is a pure-gauge solution
with zero energy
\cite{Okawa:2006vm,Fuchs:2006hw,Ellwood:2009zf,Takahashi:2007du}.
Obviously, Schnabl's solution cannot be a pure-gauge solution
since the energy of such a solution would have to be zero in
contradiction with the proven Sen's first conjecture. It is
therefore interesting to understand how the solution ceases to be
a pure-gauge in the limit $\lambda \rightarrow 1$.

It turns out that the gauge transformation (\ref{puregauge2})
becomes singular at $\lambda=1$ \cite{Ellwood:2009zf}. To see how
this can happen, let us expand the string field
$\Gamma_\lambda^{-1}$ in the $\mathcal{L}_0$ basis
\begin{align}
\label{puregauge4} \Gamma^{-1}_\lambda =&-\frac{2-\lambda }{2
(\lambda -1)}|0\rangle-\frac{\lambda }{2 (\lambda
-1)}\hat{\mathcal{B}}\tilde c_1|0\rangle+\frac{\lambda ^2-4
\lambda +2}{4 (\lambda
-1)^2}\hat{\mathcal{L}}|0\rangle-\frac{\lambda ^2}{4 (\lambda
-1)^2}\hat{\mathcal{B}}\tilde c_0|0\rangle \nonumber \\
&-\frac{\lambda ^2}{4 (\lambda -1)^2}
\hat{\mathcal{L}}\hat{\mathcal{B}}\tilde
c_1|0\rangle-\frac{\lambda ^2 (\lambda +1)}{8 (\lambda
-1)^3}\hat{\mathcal{B}}\tilde c_{-1}|0\rangle+\frac{\lambda ^3-7
\lambda ^2+6 \lambda -2}{16
   (\lambda -1)^3}\hat{\mathcal{L}}^2|0\rangle \nonumber \\
   &-\frac{\lambda ^2 (\lambda +1)}{8 (\lambda -1)^3}\hat{\mathcal{L}}\hat{\mathcal{B}}\tilde
c_0|0\rangle-\frac{\lambda ^2 (\lambda +1)}{16 (\lambda
-1)^3}\hat{\mathcal{L}}^2\hat{\mathcal{B}}\tilde
c_1|0\rangle+\cdots \, ,
\end{align}
where the dots stand for terms of higher $\mathcal{L}_0$-level.
From the expansion (\ref{puregauge4}) it is clear that the string
field $\Gamma^{-1}_\lambda$ is not well-defined at $\lambda=1$. By
this method, it is possible to show the presence of poles in the
definition of the gauge transformations, nevertheless the
expansion (\ref{puregauge4}) does not tell us much about the
interval where the parameter $\lambda$ should belong. In the
reference \cite{Takahashi:2007du} it was argued that Schnabl's
solution $\Psi^{\text{S}}_\lambda$ has a well-defined Fock space
expression if the parameter $\lambda$ belongs to the interval
$[-1,+1)$. The proof is based on the convergence properties of the
coefficients which appear when we expand $\Psi^{\text{S}}_\lambda$
in the usual Virasoro $L_0$ basis.

Coming back to the case of the gauge transformations
(\ref{regularsol1intro}) and (\ref{gauge01intro}), to determine
the interval where the parameter $\lambda$ should belong, we will
employ the same arguments developed for the case of the original
Schnabl's solution $\Psi^{\text{S}}_\lambda$. So let us first
check that if potential singularities can arise in the definition
of the gauge transformations (\ref{regularsol1intro}) and
(\ref{gauge01intro}). The place where potential singularities can
arise is in the definition of the inverse of the string field
$U_\lambda$. To see if this problem can happen, let us expand the
string field $U_\lambda^{-1}$ in the $\mathcal{L}_0$ basis
\begin{align}
\label{puregauge5} U_\lambda^{-1}
=&+|0\rangle+\Big(\frac{1}{2}-\frac{\lambda
}{4}\Big)\hat{\mathcal{L}}|0\rangle-\frac{\lambda
}{4}\hat{\mathcal{B}}\tilde c_0|0\rangle +\frac{\lambda }{4}
\hat{\mathcal{L}}\hat{\mathcal{B}}\tilde c_1|0\rangle
-\frac{\lambda ^2}{4}\hat{\mathcal{B}}\tilde c_{-1}|0\rangle
+\Big(\frac{\lambda ^2}{4}-\frac{\lambda
}{8}\Big)\hat{\mathcal{L}}\hat{\mathcal{B}}\tilde
c_0|0\rangle\nonumber \\
&+\Big(\frac{\lambda ^2}{8}-\frac{\lambda }{8}+\frac{1}{8}\Big)
\hat{\mathcal{L}}^2|0\rangle +\Big(\frac{\lambda
}{8}-\frac{\lambda
^2}{8}\Big)\hat{\mathcal{L}}^2\hat{\mathcal{B}}\tilde
c_1|0\rangle+\cdots \, .
\end{align}

From this expansion (\ref{puregauge5}), we see that the string
field $U_\lambda^{-1}$ does not have poles neither at $\lambda=1$
or $\lambda=0$. In the same way, we can also see that the
$\mathcal{L}_0$-level expansion of the regularized solution
(\ref{regularsol1intro})
\begin{align}
\label{puregauge6} \Psi_\lambda =&+\tilde c_1|0\rangle +
\frac{1}{2} \tilde c_0|0\rangle -\frac{\lambda }{2}
\hat{\mathcal{B}}\tilde c_1 \tilde c_0|0\rangle + \frac{\lambda
}{2} \hat{\mathcal{L}}\tilde c_1|0\rangle + \frac{\lambda }{2}
\tilde c_{-1}|0\rangle -\frac{\lambda }{2} \hat{\mathcal{B}}\tilde
c_1 \tilde c_{-1}|0\rangle \nonumber \\
&+\Big(\frac{1}{4}-\frac{\lambda }{4}\Big) \hat{\mathcal{L}}\tilde
c_0|0\rangle + \Big(\frac{\lambda }{4}-\frac{1}{8}\Big)
\hat{\mathcal{L}}^2 \tilde c_1|0\rangle +\frac{3 \lambda ^2}{4}
\tilde c_{-2}|0\rangle -\frac{3 \lambda ^2}{4}
\hat{\mathcal{B}}\tilde c_1 \tilde c_{-2}|0\rangle \nonumber
\\& -\frac{\lambda ^2}{4} \hat{\mathcal{B}}\tilde c_0 \tilde
c_{-1}|0\rangle +\Big(\frac{\lambda }{4}-\frac{\lambda ^2}{2}\Big)
\hat{\mathcal{L}} \tilde c_{-1}|0\rangle + \Big(\frac{\lambda
^2}{8}-\frac{\lambda }{8}+\frac{1}{16}\Big)
\hat{\mathcal{L}}^2\tilde c_0|0\rangle+\cdots \,
\end{align}
does not have poles neither at $\lambda=1$ or $\lambda=0$.

By using this method, we have just shown the absence of poles in
the definition of the gauge transformations
(\ref{regularsol1intro}) and (\ref{gauge01intro}). It remains the
question about the interval where the parameter $\lambda$ should
belong. As in the case of the original Schnabl's solution
\cite{Takahashi:2007du} to provide an answer to this question, we
need to expand the regularized solution $\Psi_\lambda$ in the
Virasoro $L_0$ basis. It turns out that the coefficients of the
$L_0$-level expansion of the regularized solution
(\ref{regularsol1intro}) are given by sums of integrals of the
form
\begin{eqnarray}
\label{integral02} \int_{0}^{\infty} e^{-t} (1+\lambda t)^m \cos^2
\big( \frac{\pi}{2} \frac{\lambda t}{1+\lambda t}\big) \tan^n
\big( \frac{\pi}{2} \frac{\lambda t}{1+\lambda t}\big)\, ,
\end{eqnarray}
for $m=2,0,-2,-4,\cdots$ and $n \in \mathbb{N}_0$. These integrals
are convergent provide that the parameter $\lambda$ belongs to the
interval $[0,+\infty)$. Therefore, as it was claimed, the gauge
transformations (\ref{regularsol1intro}) and (\ref{gauge01intro})
are well-defined for all positive values of the parameter
$\lambda$.

\section{Regularization of identity based solution in the modified cubic
superstring field theory} In this section, we extend our previous
results in order to regularize an identity based solution in the
modified cubic superstring field theory. In the superstring case,
in addition to the basic string fields $K$, $B$ and $c$, we need
to include the super-reparametrization ghost field $\gamma$ which,
in the operator representation, is given by \cite{Erler:2007xt}
\begin{eqnarray}
\label{gamma01} \gamma &\rightarrow&  U_{1}^\dag U_{1} \tilde
\gamma(0)|0\rangle \, .
\end{eqnarray}
Let us remember that in the superstring case the basic string
fields $K$, $B$, $c$ and $\gamma$ satisfy the algebraic relations
\cite{Gorbachev:2010zz,Erler:2007xt}
\begin{align}
&\{B,c\}=1\, , \;\;\;\;\;\;\; [B,K]=0 \, , \;\;\;\;\;\;\;
B^2=c^2=0
\, , \nonumber\\
\label{02eq2} \partial c = [K&,c] \, , \;\;\;\;\;\;\;
\partial \gamma  = [K,\gamma] \, , \;\;\;\;\;\;\; [c,\gamma]=0 \, ,
\;\;\;\;\;\;\; [B,\gamma]=0 \, ,
\end{align}
and have the following BRST variations
\begin{eqnarray}
\label{02eq3} QK=0 \, , \;\;\;\;\;\; QB=K \, , \;\;\;\;\;\;
Qc=cKc-\gamma^2 \, , \;\;\;\;\;\; Q\gamma=c \partial \gamma
-\frac{1}{2} \gamma
\partial c \, .
\end{eqnarray}
Employing these basic string fields, we can construct the
following identity based solution
\begin{eqnarray}
\label{02eq1} \Psi_I=(c+B \gamma^2)(1-K)
\end{eqnarray}
which formally satisfies the equation of motion $Q \Psi_I +
\Psi_I^2 =0$, where in this case $Q$ is the BRST operator of the
open Neveu-Schwarz superstring theory.

As in the bosonic case, the direct evaluation of the vacuum energy
using the identity based solution (\ref{02eq1}) brings ambiguous
result. Therefore before computing some gauge invariants, such as
the vacuum energy, first we need to regularize our identity based
solution. Using the same procedure developed in the previous
section, we show that a well behaved regularized solution
$\Psi_\lambda$ can be derived from our identity based solution
(\ref{02eq1}) by performing a gauge transformation
\begin{eqnarray}
\label{regularsupero1}\Psi_\lambda &=& U_\lambda (Q + \Psi_I) U_\lambda^{-1} \nonumber \\
&=& \Big[\lambda cBK +1\Big] \Big( Q + (c+B \gamma^2)(1-K)
\Big)  \Big[1-\lambda cBK\frac{1}{1+\lambda K}\Big]\nonumber \\
&=& (c + \lambda cKBc+B\gamma^2)\frac{1+(\lambda-1)K}{1+\lambda K}
\, .
\end{eqnarray}

Note that this regularized solution interpolates between the
identity based solution (\ref{02eq1}) which corresponds to the
case $\lambda\rightarrow 0$, and the Gorbachev's solution
\cite{Gorbachev:2010zz} which corresponds to the case
$\lambda\rightarrow 1$. In the next subsection we are going to
evaluate the kinetic term for the regularized solution, and it
will be shown that its value does not depend on the parameter
$\lambda$.

\subsection{The kinetic term}
In this subsection, we are going to evaluate the kinetic term of
the modified cubic superstring field theory action for the
regularized solution $\Psi_\lambda$
\begin{eqnarray}
\label{kinetic001II} \langle \langle \Psi_\lambda,Q \Psi_\lambda
\rangle \rangle\,.
\end{eqnarray}
The inner product $\langle \langle \cdot , \cdot \rangle \rangle$
is the standard BPZ inner product with the difference that we must
insert the operator $Y_{-2}$ at the open string midpoint. The
operator $Y_{-2}$ can be written as the product of two inverse
picture changing operators $Y_{-2}=Y(i)Y(-i)$, where
$Y(z)=-\partial \xi e^{-2 \phi} c(z)$.

In order to simplify the computations, let us write the
regularized solution (\ref{regularsupero1}) as an expression
containing two terms
\begin{eqnarray}
\label{regularsolz1} \Psi_\lambda &=& \Psi_1 + \Psi_2 \, , \\
\label{regularsolz2} \Psi_1&=& c \frac{1+(\lambda-1)K}{1+\lambda
K} \, , \\
\label{regularsolz3} \Psi_2&=&  (\lambda cBKc + B \gamma^2)
\frac{1+(\lambda-1)K}{1+\lambda K} \, .
\end{eqnarray}

Replacing equations (\ref{regularsolz1})$-$(\ref{regularsolz3})
into the expression for the kinetic term (\ref{kinetic001II}), we
obtain
\begin{eqnarray}
\label{kinetic001z} \langle \langle \Psi_\lambda,Q \Psi_\lambda
\rangle \rangle =\langle \langle \Psi_1, Q \Psi_1 \rangle \rangle
+ 2\langle \langle \Psi_1,Q \Psi_2 \rangle\rangle + \langle
\langle \Psi_2,Q \Psi_2 \rangle \rangle \, ,
\end{eqnarray}
each term on the RHS of (\ref{kinetic001z}) is given by
\begin{eqnarray}
\label{kinetic003z1} \langle \langle \Psi_1, Q \Psi_1 \rangle
\rangle &=& -\int_0^{\infty}\int_0^{\infty} dt_1dt_2 e^{-t_1-t_2}
\mathcal{D}_1 \mathcal{D}_2 \langle \langle c \Omega^{\lambda t_1}
\gamma^2 \Omega^{\lambda t_2}  \rangle \rangle  \, , \\
\label{kinetic003z2} \langle \langle \Psi_1,Q \Psi_2
\rangle\rangle &=& 2(1-\lambda)\int_0^{\infty}\int_0^{\infty}
dt_1dt_2 e^{-t_1-t_2} \mathcal{D}_1 \mathcal{D}_2 \langle \langle
c \Omega^{\lambda t_1} cBK\gamma^2 \Omega^{\lambda t_2} \rangle
\rangle  \, ,
\\
\label{kinetic003z3} \langle \langle \Psi_2,Q \Psi_2 \rangle
\rangle &=& 2\lambda (1-\lambda)\int_0^{\infty}\int_0^{\infty}
dt_1dt_2 e^{-t_1-t_2} \mathcal{D}_1 \mathcal{D}_2  \langle \langle
cBKc \Omega^{\lambda t_1} cBK\gamma^2 \Omega^{\lambda
t_2}  \rangle \rangle \nonumber \, . \\
\end{eqnarray}

The correlators $\langle \langle c \Omega^{\lambda t_1} \gamma^2
\Omega^{\lambda t_2}  \rangle \rangle $, $\langle \langle c
\Omega^{\lambda t_1} cBK\gamma^2 \Omega^{\lambda t_2} \rangle
\rangle$ and $\langle \langle cBKc \Omega^{\lambda t_1}
cBK\gamma^2 \Omega^{\lambda t_2}  \rangle \rangle$ can be computed
using the methods given in the appendix. Plugging the expression
for these correlators into equations
(\ref{kinetic003z1})$-$(\ref{kinetic003z3}), applying the change
of variables as in \cite{Erler:2009uj} $t_1 \rightarrow u v$, $t_2
\rightarrow  u (1 - v)$ and performing the $v$ integral, we obtain
from (\ref{kinetic001z})
\begin{eqnarray}
\label{kinetic003zz} \langle \langle  \Psi_\lambda,Q \Psi_\lambda
\rangle \rangle &=& -\frac{1}{2 \pi^2} \int_0^{\infty} du e^{-u}
\big[ 6 (\lambda -1)^2 u -6 (\lambda -1) \lambda u^2 +\lambda ^2
u^3\big] \nonumber
\\
&=&-\frac{3}{\pi ^2} \, .
\end{eqnarray}
Therefore, as in the bosonic case, the value of the kinetic term
does not depend on the parameter $\lambda$. At this stage, we can
safely take the limit $\lambda \rightarrow 0$ which corresponds to
the identity based solution.

If we assume the validity of the equation of motion when
contracted with the solution itself, it is clear that the value of
the vacuum energy can be correctly reproduced. Nevertheless it is
important to test whether or not the equation of motion is
satisfied when it is contracted with the regularized solution
itself. To prove the correctness of this statement, it is
necessary to compute explicitly the cubic term of the modified
cubic superstring field theory action.

\subsection{The cubic term}
In this subsection, we are going to evaluate the cubic term of the
modified cubic superstring field theory action for the regularized
solution
\begin{eqnarray}
\label{cubic001z} \langle \langle  \Psi_\lambda,\Psi_\lambda *
\Psi_\lambda \rangle \rangle \,.
\end{eqnarray}

Since the regularized solution (\ref{regularsupero1}) can be
written as an expression containing two terms (\ref{regularsolz2})
and (\ref{regularsolz3}), the calculation of the cubic term
(\ref{cubic001z}) can be reduced to the evaluation of the
following correlators
\begin{eqnarray}
\label{cubic002z} \langle \langle \Psi_\lambda, \Psi_\lambda *
\Psi_\lambda \rangle \rangle = \langle \langle \Psi_1, \Psi_1 *
\Psi_1\rangle \rangle+ 3 \langle \langle \Psi_2, \Psi_1 *
\Psi_1\rangle \rangle+ 3 \langle \langle \Psi_2, \Psi_2 *
\Psi_1\rangle \rangle + \langle \langle \Psi_2, \Psi_2
* \Psi_2\rangle \rangle \,
, \nonumber \\
\end{eqnarray}
each term on the RHS of (\ref{cubic002z}) is given by
\begin{eqnarray}
\label{cubic003z1} \langle\langle \Psi_2, \Psi_1 *
\Psi_1\rangle\rangle &=&
\int_0^{\infty}\int_0^{\infty}\int_0^{\infty} dt_1dt_2dt_3
e^{-t_1-t_2-t_3} \mathcal{D}_1 \mathcal{D}_2 \mathcal{D}_3 \langle
\langle B \gamma^2 \Omega^{\lambda t_1} c \Omega^{\lambda t_2} c
\Omega^{\lambda
t_3} \rangle \rangle  \, , \\
\label{cubic003z2} \langle \langle \Psi_2, \Psi_2 * \Psi_1 \rangle
\rangle &=& 2 \lambda
\int_0^{\infty}\int_0^{\infty}\int_0^{\infty} dt_1dt_2dt_3
e^{-t_1-t_2-t_3} \mathcal{D}_1 \mathcal{D}_2 \mathcal{D}_3
\langle\langle B \gamma^2  \Omega^{\lambda t_1} cBKc
\Omega^{\lambda t_2} c \Omega^{\lambda t_3} \rangle \rangle
\nonumber  \, ,
\\\\
\label{cubic003z3} \langle\langle \Psi_2, \Psi_2 * \Psi_2
\rangle\rangle &=& 3 \lambda^2
\int_0^{\infty}\int_0^{\infty}\int_0^{\infty} dt_1dt_2dt_3
e^{-t_1-t_2-t_3} \mathcal{D}_1 \mathcal{D}_2 \mathcal{D}_3
\langle\langle B\gamma^2 \Omega^{\lambda t_1} cBKc \Omega^{\lambda
t_2} cBKc \Omega^{\lambda
t_3} \rangle\rangle \nonumber \, . \\
\end{eqnarray}
For the correlators to be nonzero, they must have a
$\phi$-momentum equal to $-2$, since the picture changing operator
has a $\phi$-momentum equal to $-4$, the first term on the RHS of
(\ref{cubic002z}) gives vanishing result.

The expression for the correlators $\langle \langle B \gamma^2
\Omega^{\lambda t_1} c \Omega^{\lambda t_2} c \Omega^{\lambda t_3}
\rangle \rangle$, $\langle\langle B \gamma^2 \Omega^{\lambda t_1}
cBKc \Omega^{\lambda t_2} c \Omega^{\lambda t_3} \rangle \rangle$
and $\langle\langle B\gamma^2 \Omega^{\lambda t_1} cBKc
\Omega^{\lambda t_2} cBKc \Omega^{\lambda t_3} \rangle\rangle$ can
be derived using the techniques developed in the appendix.
Plugging the expression for these correlators into equations
(\ref{cubic003z1})$-$(\ref{cubic003z3}), applying the change of
variables as in \cite{Zeze:2010jv} $t_1 \rightarrow u v_1$, $t_2
\rightarrow  u v_2$, $t_3 \rightarrow  u (1 - v_1 - v_2)$ and
performing the $v_1,v_2$ integral, we obtain from
(\ref{cubic002z})
\begin{align}
\label{cubic0zz1} \langle \langle  \Psi_\lambda, \Psi_\lambda *
\Psi_\lambda \rangle\rangle &= \frac{1}{4 \pi^2} \int_0^{\infty}
du e^{-u} \big[ 6 (\lambda -1) (2 \lambda -1)u^2 -2 \lambda  (4
\lambda -3) u^3+\lambda ^2 u^4 \big] \nonumber
\\
&=\frac{3}{\pi ^2} \, .
\end{align}

As it was expected, we see that the value of the cubic term does
not depend on the parameter $\lambda$. This result
(\ref{cubic0zz1}) proves the statement that the equation of
motion, in the modified cubic superstring field theory, is
satisfied when it is contracted with the regularized solution
itself.

\subsection{$\mathcal{L}_0$ level expansion}

Although we have the expected result for the value of the kinetic
(\ref{kinetic003zz}) and cubic term (\ref{cubic0zz1}), we would
like to confirm our calculation by using the $\mathcal{L}_0$ level
expansion of the solution. In order to evaluate the kinetic and
cubic term of the modified cubic superstring field theory action
in the $\mathcal{L}_0$ level expansion scheme, let us write the
regularized solution (\ref{regularsupero1}) in terms of
$\mathcal{L}_0$ eigenstates
\begin{align}
\label{L0Super} \Psi_\lambda =& \sum_{n,p} f_{n,p}(\lambda) ({\cal
L}_0 + {\cal L}_0^\dagger)^n \tilde c_p |0\rangle  + \sum_{n,p,q}
f_{n,p,q}(\lambda) ({\cal B}_0 + {\cal B}_0^\dagger) ({\cal L}_0 +
{\cal
L}_0^\dagger)^n \tilde c_p \tilde c_q  |0\rangle \nonumber \\
&+ \sum_{n,t,u} g_{n,t,u}(\lambda) ({\cal B}_0 + {\cal
B}_0^\dagger) ({\cal L}_0 + {\cal L}_0^\dagger)^n \tilde \gamma_t
\tilde \gamma_u  |0\rangle \, ,
\\
f_{n, p}(\lambda) =& 2^{-n} \frac{(\lambda -1)}{\lambda}
\left(\frac{1}{
   n!}+\frac{\lambda}{(n-1)!}\right) \delta _{1,p} \nonumber \\
  &+\frac{2^{-n+p-1}}{\lambda } \sum_{k=0}^{n} \frac{(-1)^{n-k} \lambda ^{n-k-p+1} (n-k-p+1)!}{k!
   (n-k)!} \nonumber \\&+ 2^{-n+p-1} \sum_{k=0}^{n-1} \frac{(-1)^{n-k-1} \lambda ^{n-k-p} (n-k-p)!}{k! (n-k-1)!} \, , \\
f_{n, p, q}(\lambda) =& \frac{2^{-n-2} (1-\lambda )}{n!}
\big(\delta _{0,q} \delta _{1,p}-\delta _{0,p} \delta _{1,q} \big)
\nonumber \\
&+2^{-n+p+q-3} (q-p) \sum_{k=0}^{n}\frac{(-1)^{n-k} \lambda
^{n-k-p-q+1} (n-k-p-q+1)!}{k! (n-k)!} \, ,
\\
g_{n, t, u}(\lambda) =&2^{-n+t+u-2} \sum_{k=0}^{n}\frac{(-1)^{n-k}
\lambda ^{n-k-t-u+1} (n-k-t-u+1)!}{k! (n-k)!}\nonumber \\
&+(\lambda-1)2^{-n+t+u-2} \sum_{k=0}^{n-1}\frac{(-1)^{n-k-1}
\lambda ^{n-k-t-u} (n-k-t-u)!}{k! (n-k-1)!}\nonumber \\
&+(1-\lambda)(1-t-u)2^{-n+t+u-2} \sum_{k=0}^{n}\frac{(-1)^{n-k}
\lambda ^{n-k-t-u} (n-k-t-u)!}{k! (n-k)!} \, .
\end{align}

Next we follow the same steps developed in the bosonic case.
Replacing the $\mathcal{L}_0$ level expansion of the regularized
solution $\Psi_\lambda$ (\ref{L0Super}) with $z^{\mathcal{L}_0}
\Psi_\lambda$ and plugging it into the kinetic term, we arrive to
\begin{align}
\label{kineticPaSuper} \langle \langle \Psi_\lambda,
z^{\mathcal{L}_0^{\dagger}} Q z^{\mathcal{L}_0} \Psi_\lambda
\rangle \rangle = -\frac{2}{\pi ^2 z^2} + \left(\frac{2 \lambda
}{\pi ^2}-\frac{2}{\pi ^2}\right)\frac{1}{z} + \left(\frac{1}{\pi
^2}-\frac{2 \lambda }{\pi ^2}\right) \, .
\end{align}
Since equation (\ref{kineticPaSuper}) has a finite number of
terms, there is no need for Pad\'{e} approximants, therefore
setting $z=1$ from equation (\ref{kineticPaSuper}) we obtain
\begin{align}
\label{kineticPaSuper001} \langle \langle \Psi_\lambda,
 Q  \Psi_\lambda
\rangle \rangle = -\frac{2}{\pi ^2} + \left(\frac{2 \lambda }{\pi
^2}-\frac{2}{\pi ^2}\right) + \left(\frac{1}{\pi ^2}-\frac{2
\lambda }{\pi ^2}\right) = - \frac{3}{\pi^2}\, .
\end{align}

As we have noticed, in the superstring case the power series
expansion of the kinetic term (\ref{kineticPaSuper}) has a finite
number of terms, this result is in contrast to the bosonic case,
where the series has an infinite number of terms
(\ref{kineticPa}). The reason for this result is due to the fact
that the correlation functions in the superstring case are much
simpler than the bosonic ones \footnote{Let us point out that a
similar result was found in \cite{Erler:2007xt} for the original
Erler's solution.}. As it is summarized in appendix A, in the
bosonic case the expressions for the correlators are given in
terms of trigonometrical functions, while in the superstring case
the expressions for the correlators are given in terms of
polynomials.

To evaluate the cubic term using its formal power series expansion
in $z$, we follow the same steps developed in the case of the
kinetic term. Plugging the $\mathcal{L}_0$ level expansion of the
regularized solution (\ref{L0Super}) into the cubic term, we get
\begin{align}
\label{expansion1Super} \langle \langle \Psi_\lambda ,z^{{\cal
L}_0^\dag}(z^{{\cal L}_0}\Psi_\lambda)*(z^{{\cal
L}_0}\Psi_\lambda) \rangle \rangle  = \frac{9}{2 \pi ^2 z^2}
-\frac{3 \lambda }{\pi ^2 z} + \left(\frac{3 \lambda }{\pi
^2}-\frac{3}{2 \pi ^2} \right)\, .
\end{align}
Since this last equation (\ref{expansion1Super}) has a finite
number of terms, as in the case of the kinetic term, there is no
need for Pad\'{e} approximants, therefore setting $z=1$ from
equation (\ref{expansion1Super}) we obtain
\begin{align}
\label{expansion1Super001} \langle \langle \Psi_\lambda
,\Psi_\lambda*\Psi_\lambda \rangle  \rangle = \frac{9}{2 \pi ^2 }
-\frac{3 \lambda }{\pi ^2 } + \left(\frac{3 \lambda }{\pi
^2}-\frac{3}{2 \pi ^2} \right)= \frac{3}{\pi ^2}\, .
\end{align}

\section{Summary and discussion}

We have shown that our recently proposed identity based solutions
\cite{Arroyo:2010fq}, in open bosonic string field theory as well
as in the modified cubic superstring field theory, can be
consistently regularized. By consistent we mean that the resulting
regularized solution brings the right value for the kinetic and
cubic term of the string field theory action.

We have proved the correctness of the above statement by employing
two different means: straightforward analytical computations and
by using the $\mathcal{L}_0$ level expansion of the solution. It
turns out that, in the bosonic case, employing the $\mathcal{L}_0$
level expansion scheme the use of Pad\'{e} approximants was
needed, while in the superstring case the expected value for the
kinetic and cubic term was derived without the use of Pad\'{e}
approximants. As a consequence of these results, we proved that
the assumption of the validity of the equation of motion when
contracted with the regularized solution itself was nevertheless
correct.

It would be important to extend this analysis to the case of
Berkovits WZW-type superstring field theory
\cite{Berkovits:1995ab}, since this theory has a non-polynomial
action, the issue for finding the tachyon vacuum solution and the
computation of the value of the D-brane tension seems to be highly
cumbersome. Nevertheless, we hope that the ideas developed in this
paper should be very useful in order to solve this challenging
puzzle.

One more significant application of the techniques established in
this paper, as discussed in \cite{Zeze:2010jv}, should be the
extension of the subalgebra generated by the basic string fields
$K$, $B$, $c$ and $\gamma$ in order to analyze identity based
solutions in more general string field configurations
\cite{Garousi:2008ge,Garousi:2007fk}, such that multiple D-branes,
marginal deformations, lump solutions as well as time dependent
solutions.

\section*{Acknowledgements}
I would like to thank Ted Erler, Michael Kroyter and Syoji Zeze
for useful discussions. This work is supported by FAPESP grant
2010/05990-0.

\appendix
\setcounter{equation}{0}

\def\thesection{\Alph{section}}
\renewcommand{\theequation}{\Alph{section}.\arabic{equation}}

\section{Correlation functions} In this appendix we provide the
details related to the derivation of the correlators used in
equations (\ref{kinetic002}),
(\ref{cubic003x1})$-$(\ref{cubic003x3}),
(\ref{kinetic003z1})$-$(\ref{kinetic003z3}) and
(\ref{cubic003z1})$-$(\ref{cubic003z3}). Let us start with the
correlators
\begin{align}
\label{ape01} \langle c \Omega^{\lambda t_1} cKc \Omega^{\lambda
t_2} \rangle &= -\partial_{s_1} \Big[\big\langle c(s_1+t_1 \lambda
+t_2 \lambda) c(s_1+t_2 \lambda)c(t_2 \lambda)\big\rangle_{s_1+t_1
\lambda +t_2 \lambda}
\Big]\Big\arrowvert_{s_1=0} \, , \\
\label{ape02}  \langle c \Omega^{\lambda t_1} c \Omega^{\lambda
t_2} c \Omega^{\lambda t_3} \rangle &= \big\langle c(t_1 \lambda
+t_2 \lambda +t_3 \lambda ) c(t_2 \lambda
   +t_3 \lambda)c(t_3 \lambda )\big\rangle_{t_1 \lambda +t_2 \lambda +t_3
   \lambda}  \, ,
\end{align}
where the expression for the correlator $\langle
c(x_1)c(x_2)c(x_3)\rangle_{L}$ is given by
\begin{eqnarray}
\label{ape03}\langle c(x_1)c(x_2)c(x_3)\rangle_{L} =
\frac{L^3}{\pi ^3} \sin \left(\frac{\pi
\left(x_1-x_2\right)}{L}\right) \sin
   \left(\frac{\pi  \left(x_1-x_3\right)}{L}\right) \sin \left(\frac{\pi
   \left(x_2-x_3\right)}{L}\right) \, .
\end{eqnarray}

For the computation of the firsts two correlators (\ref{ape01})
and (\ref{ape02}), the correlator (\ref{ape03}) is all we need,
for instance using (\ref{ape03}) from (\ref{ape01}) we obtain
\begin{eqnarray}
\label{ape04} \langle c \Omega^{\lambda t_1} cKc \Omega^{\lambda
t_2} \rangle = -\frac{\lambda ^2
   \left(t_1+t_2\right)^2}{\pi ^2} \sin ^2\left(\frac{\pi
t_1}{t_1+t_2}\right) \, .
\end{eqnarray}

The next two correlators, which were used in the computation of
the cubic term of the open bosonic string field theory action, are
given by
\begin{align}
\label{cubic003x2corr}  \langle cBKc \Omega^{\lambda t_1} cBKc
\Omega^{\lambda t_2} c \Omega^{\lambda t_3} \rangle &=
\partial_{s_1} \partial_{s_2}\Big[\big \langle c(\alpha_1)Bc(\alpha_2) c(\alpha_3)Bc(\alpha_4)  c(\alpha_5)  \big \rangle_{(t_1  +t_2
+t_3)
   \lambda+s_1+s_2}\Big]\Big\arrowvert_{s_1=0,s_2=0} \, ,
\\
\label{cubic003x3corr} \langle cBKc \Omega^{\lambda t_1} c
\Omega^{\lambda t_2} c \Omega^{\lambda t_3} \rangle
&=-\partial_{s_1} \Big[\big \langle c(\beta_1)Bc(\beta_2)
c(\beta_3) c(\beta_4) \big \rangle_{(t_1 +t_2  +t_3)
   \lambda+s_1}\Big]\Big\arrowvert_{s_1=0} \, ,
\end{align}
where
\begin{align}
\alpha_1&=(t_1+t_2+t_3)\lambda+s_1+s_2 \; , \;\;\;
\alpha_2=(t_1+t_2+t_3)\lambda+s_2 \; , \nonumber \\
\alpha_3&=(t_2+t_3)\lambda+s_2 \,\; , \;\;\;\;\;\;\;\;\,
\alpha_4=(t_2+t_3)\lambda \; , \;\;\;\;\;\;\; \alpha_5=t_3 \lambda
\; , \nonumber
\\
\beta_1&=(t_1+t_2+t_3)\lambda+s_1\; , \;\;\,
\beta_2=(t_1+t_2+t_3)\lambda \; , \nonumber \\
\beta_3&=(t_2+t_3)\lambda\; , \;\;\;\;\;\;\;\;\;\;\;\;\;\;\;\;\;
\beta_4=t_3\lambda \, .
\end{align}

The correlators $\langle c(\alpha_1)Bc(\alpha_2)
c(\alpha_3)Bc(\alpha_4)  c(\alpha_5)  \rangle_{L}$ and $\langle
c(\beta_1)Bc(\beta_2) c(\beta_3) c(\beta_4)\rangle_{L}$ can be
computed using the following correlator \cite{Okawa:2006vm}
\begin{align}
\label{corrB} \langle B c(x_1)c(x_2) c(x_3) c(x_4)\rangle_{L} &=
\frac{x_1}{L}  \langle c(x_2)c(x_3)c(x_4)\rangle_{L} -
\frac{x_2}{L}  \langle c(x_1)c(x_3)c(x_4)\rangle_{L} \nonumber
\\&+ \frac{x_3}{L}  \langle c(x_1)c(x_2)c(x_4)\rangle_{L}
-\frac{x_4}{L} \langle c(x_1)c(x_2)c(x_3)\rangle_{L} \, .
\end{align}

In the case of the modified cubic superstring field theory, the
expressions for the correlators are much easier than the ones
given in the bosonic case \cite{Erler:2007xt}
\begin{eqnarray}
\label{kinetic003z1corr}  \langle \langle c \Omega^{\lambda t_1}
\gamma^2 \Omega^{\lambda t_2}  \rangle \rangle&=&\frac{\lambda ^2 \left(t_1+t_2\right)^2}{2 \pi ^2}  \, , \\
\label{kinetic003z2corr}  \langle \langle c \Omega^{\lambda t_1}
cBK\gamma^2 \Omega^{\lambda t_2} \rangle \rangle &=&-\frac{\lambda
t_1}{2 \pi ^2} \, ,
\\
\label{kinetic003z3corr}  \langle \langle cBKc \Omega^{\lambda
t_1} cBK\gamma^2 \Omega^{\lambda t_2}  \rangle \rangle &=&0\, ,
\\
\label{cubic003z1corr}  \langle \langle B \gamma^2 \Omega^{\lambda
t_1} c \Omega^{\lambda t_2} c \Omega^{\lambda
t_3} \rangle \rangle &=& \frac{\lambda ^2 t_2 \left(t_1+t_2+t_3\right)}{2 \pi ^2} \, , \\
\label{cubic003z2corr} \langle\langle B \gamma^2  \Omega^{\lambda
t_1} cBKc \Omega^{\lambda t_2} c \Omega^{\lambda t_3} \rangle
\rangle &=& -\frac{\lambda  t_2}{2 \pi ^2}   \, ,
\\
\label{cubic003z3corr} \langle\langle B\gamma^2 \Omega^{\lambda
t_1} cBKc \Omega^{\lambda t_2} cBKc \Omega^{\lambda t_3}
\rangle\rangle &=&0\, .
\end{eqnarray}
To derive these correlators, we have used the following two basic
correlators \footnote{These correlation functions has been
computed using the normalization: $\langle \xi(x) c \partial c
\partial^2 c (y) e^{-2 \phi(z)} \rangle=2.$}
\begin{eqnarray}
\label{corresuper01} \langle\langle c(x_1) \gamma^2(x_2)
\rangle\rangle_L
&=&\frac{L^2}{2 \pi ^2} \, ,\\
\langle\langle B c(x_1)c(x_2) \gamma^2(x_3) \rangle\rangle_L &=&
\frac{L (x_1-x_2)}{2 \pi ^2} \, .
\end{eqnarray}

\section{Pad\'{e} approximant computations}
In this appendix, as a pedagogical illustration of the numerical
method based on Pad\'{e} approximants, we are going to compute in
detail the normalized value of the kinetic term
$\frac{\pi^2}{3}\langle \Psi_\lambda, z^{\mathcal{L}_0^{\dagger}}
Q z^{\mathcal{L}_0} \Psi_\lambda \rangle$ at order $n = 2$. At
this order we need to consider terms in the series expansion of
the kinetic term (\ref{kineticPa}) up to quadratic order in $z$,
namely
\begin{eqnarray}
\label{order2ki} -\frac{4}{\pi ^2 z^2}+\frac{(\pi ^2-4) (2 \lambda
-1)}{2 \pi ^2} + \frac{\pi ^2 (\lambda ^2-2 \lambda^3 )}{8}  z^2
\, .
\end{eqnarray}

Using the numerical method based on Pad\'{e} approximants, first
we express (\ref{order2ki}) as the following rational function
\begin{eqnarray}
\label{ss2} P^2_{2+2}(z)=\frac{1}{z^2} \Big[\frac{a_0+a_1z+a_2z^2
}{1+b_1z+b_2z^2} \Big]\, .
\end{eqnarray}
Expanding the right hand side of (\ref{ss2}) around $z=0$, we get
up to quadratic order in $z$
\begin{eqnarray}
\label{ss3} P^2_{2+2}(z)&=&\frac{a_0}{z^2} +
\frac{a_1-a_0b_1}{z}+(a_2 - a_1b_1 + a_0 b_1^2 - a_0
b_2) \nonumber \\
&+&( a_1b_1^2 -a_2b_1 - a_0b_1^3 - a_1b_2 +
      2a_0b_1b_2) z \nonumber \\
&+&(a_2b_1^2 - a_1 b_1^3 + a_0b_1^4 - a_2 b_2 +
          2 a_1b_1 b_2 - 3 a_0 b_1^2 b_2 +
          a_0b_2^2) z^2 \, .
\end{eqnarray}

Equating the coefficients of $z^{-2}$, $z^{-1}$, $z^{0}$, $z^{1}$,
$z^{2}$ in equations (\ref{order2ki}) and (\ref{ss3}), we get a
system of five algebraic equations for the unknown coefficients
$a_0$, $a_1$, $a_2$, $b_1$ and $b_2$. Solving these equations we
get
\begin{eqnarray}
\label{aaa1}a_0&=&-\frac{4}{\pi ^2} \, , \\
\label{aaa2}a_1&=&0 \, , \\
\label{aaa3}a_2&=& \frac{\pi ^2 (8-16 \lambda )+32 \lambda +\pi ^4
   \left(-2 \lambda ^2+2 \lambda -1\right)-16}{2 \pi
   ^2 \left(-4+\pi ^2\right)} \, , \\
\label{aaa4}b_1&=& 0 \, , \\
\label{aaa5}b_2&=& \frac{\pi ^4 \lambda ^2}{4 \left(-4+\pi
^2\right)} \, .
\end{eqnarray}

Replacing the value of the coefficients (\ref{aaa1}),
(\ref{aaa2}), (\ref{aaa3}), (\ref{aaa4}) and (\ref{aaa5}) into the
definition of $P^2_{2+2}(z)$ (\ref{ss2}), and evaluating this at
$z=1$, we get the following normalized value for the kinetic term,
\begin{eqnarray}
\label{n2kine} \frac{\pi^2}{3} P^2_{2+2}(z=1) = \frac{-32 \pi ^2
\lambda +64 \lambda +\pi ^4 \left(-4
   \lambda ^2+4 \lambda -2\right)+32}{3 \left(\pi ^4
   \lambda ^2+4 \pi ^2-16\right)} \, .
\end{eqnarray}

Now we are ready to evaluate this expression (\ref{n2kine}) for
some particular values of $\lambda$. Our first interest is the
case $\lambda \rightarrow 0$ which corresponds to the identity
based solution, where we get
\begin{eqnarray}
\label{n2kine0} \frac{\pi^2}{3} P^2_{2+2}(z=1;\lambda \rightarrow
0) = -2.311600733514 \, ,
\end{eqnarray}
while in the case of $\lambda \rightarrow 1$ which corresponds to
the Erler-Schnabl's solution, we obtain
\begin{eqnarray}
\label{n2kine1} \frac{\pi^2}{3} P^2_{2+2}(z=1;\lambda \rightarrow
1) = -1.143337106188 \, .
\end{eqnarray}

\newpage

\end{document}